\documentclass[mathleft
]{an}
\usepackage{graphicx}
\usepackage{times}
\begin{document}

\Pagespan{789}{}
\Yearpublication{2006}%
\Yearsubmission{2005}%
\Month{11}%
\Volume{999}%
\Issue{88}%

\title{Accuracy of the numerical computation of solar g modes}

\author{A. Moya\inst{1}\fnmsep\thanks{Corresponding author:
  \email{amoya@cab.inta-csic.es}\newline}, S. Mathur\inst{2} \and R. A. Garc\'ia\inst{3}
}

\titlerunning{Accuracy of the numerical computation of solar g-modes} \authorrunning{A. Moya et al.}  \institute{ Departamento de
  Astrof\'{\i}sica, Laboratorio de Astrof\'{\i}sica Estelar y
  Exoplanetas, LAEX-CAB (INTA-CSIC), PO BOX 78, 28691 Villanueva de la
  Ca\~nada, Madrid, Spain
   \and
   High Altitude Observatory, NCAR, P.O. Box 3000, Boulder, CO 80307, USA
 \and
  Laboratoire AIM, CEA/DSM-CNRS - U. Paris Diderot - IRFU/SAp, CEA-Saclay, 91191 Gif-sur-Yvette,
  Cedex, France }
\publonline{later}


\abstract{From the recent work of the Evolution and Seismic Tools
  Activity (ESTA, Monteiro et al. 2006; Lebreton et al. 2008), whose
  Task 2 is devoted to compare pulsational frequencies computed using
  most of the pulsational codes available in the asteroseismic
  community, the dependence of the theoretical frequencies with
  non-physical choices is now quite well fixed. To ensure that the
  accuracy of the computed frequencies is of the same order of
  magnitude or better than the observational errors, some requirements
  in the equilibrium models and the numerical resolutions of the
  pulsational equations must be followed. In particular, we have
  verified the numerical accuracy obtained with the Saclay seismic
  model, which is used to study the solar g-mode region (60 to
  140$\mu$Hz). We have compared the results coming from the Aarhus
  adiabatic pulsation code (ADIPLS), with the frequencies computed
  with the Granada Code (GraCo) taking into account several possible
  choices. We have concluded that the present equilibrium models and
  the use of the Richardson extrapolation ensure an accuracy of the
  order of $0.01 \mu Hz$ in the determination of the frequencies,
  which is quite enough for our purposes.}

\maketitle

\section{Introduction}

The interior of the Sun has been very well studied thanks to the
information provided by pressure-driven modes ($p$ modes). In the case
of the dynamics of the solar interior, due to the very small number of
non-radial $p$ modes penetrating inside the core, neither the rotation
profile (e.g., Chaplin et al. 1999; Thomson et al. 2003; Garc\'ia et
al. 2001, 2004, 2008c) nor the dynamical processes (e.g., Mathis \&
Zahn 2004, 2005) are well constrained inside this region.

On the other hand, gravity ($g$) modes would give us complete access
to the solar core, in particular, to its dynamics (e.g., Mathur et
al. 2008; Mathur et al. 2010).

Gravity modes have been searched for a long time, almost since the
beginning of helioseismology (e.g. Hill et al. 1991; Pall\'e
1991). But there is currently no undisputed detection of individual
$g$ modes for the Sun (Appourchaux et al 2010). However, some peaks
(e.g. Gabriel et al. 2002; Jim\'enez \& Garc\'ia 2009) and groups of
peaks (Turck-Chi\`eze et al. 2004; Garc\'ia et al. 2008a) have been
considered as reliable $g$-mode candidates as they are above than
90$\%$ confidence level and they present several of their expected
properties. Moreover, to increase the probability of detection,
Garc\'ia et al. (2007, 2008b) searched for the global signature of
such modes instead of looking for individual $g$ modes. They have
found the signature of the asymptotic-dipole $g$ modes with more than
99.99$\%$ confidence level. The detailed study of this asymptotic
periodicity revealed a higher rotation rate in the core than in the
rest of the radiative region and a better agreement with solar models
(Garc\'ia et al. 2008b) computed with old-surface abundances
(Grevesse, Noels, \& Sauval 1993) compared to the new ones (Asplund,
Grevesse \& Sauval 2005).  However, it was not possible to identify
the sequence of individual peaks generating the detected signal
because of the very small signal-to-noise ratio. Thus, to go further
it is necessary to use theoretical $g$-mode predictions to guide our
search (Broomhall et al. 2007; Garc\'\i a 2010). For this purpose, we
need to know the limits of the modeled physical processes and
quantities as well as the internal numerical errors of the codes used
to compute the predicted frequencies.

The accuracy of the present solar models has already been studied by
Mathur et al. (2007) and Zaatri et al. (2007). They showed that models
with different physical inputs and fixed surface abundances present
differences in the frequencies of the $g$ modes that are below 1
$\mu$Hz in the range [60, 140] $\mu$Hz.

In the present work we study the numerical errors introduced by the
approaches followed by the oscillation codes used to compute the
$g$-mode frequencies of the Sun. It is a direct application of the
study done in the ESTA group (whose Task 2 is devoted to the
pulsational code's comparison, see Moya et al. 2008) to the solar case
and the calculation of the $g$-mode frequencies. This comparison makes
it possible to fix the global uncertainties of the numerical schemes
used (Moya et al. 2010a).


\section{Modeling the Sun Interior: Computing p and $g$ Modes}

\subsection{Solar Model}

We have computed one solar model, which is based on the Seismic model
developed by the Saclay team (Turck-Chi\`e\-ze et al. 2001; Couvidat
et al. 2003). It is a 1-D model computed with the so called {\it Code
  d'Evolution Stellaire Adaptatif et Modulaire} (CESAM, Morel
1997). This solar model was tuned to better match helioseismic
observations (e.g. the sound speed profile), especially in the
radiative region. This model was also used to more accurately predict
the neutrino fluxes. It is calibrated in terms of surface metallicity,
luminosity, and radius at the age of 4.6~Gyrs with an accuracy of
$10^{-5}$.

As one of the main aims of this study consists of calculating the
frequencies of $g$ modes, and these modes are mainly confined in the
inner regions of the Sun, we have computed the model using the highest
resolution in the core of the Sun (below 0.5~$R_\odot$).

\subsection{Pulsation Codes Brief Description: Aarhus and GraCo}

To calculate the $p$- and $g$-mode frequencies we have used two
oscillation codes: the Granada Code (GraCo) and the Aarhus adiabatic
pulsation code (ADIPLS), both being part of the ESTA group.

GraCo (Moya et al. 2004; Moya \& Garrido 2008) is a non-radial
non-adiabatic linear pulsational code using a second-order integration
scheme to solve the set of differential equations. The code also
provides frequencies in the adiabatic approximation.  This code has
been used as a reference code for the ESTA study. Thus all of
the possible numerical schemes and algorithms in the literature have
been implemented. Therefore, with this code, it is possible to study the
numerical accuracy of the frequencies obtained with any singular
equilibrium model. In adition, this code has been used to study
several $g$-mode pulsators (Rodr\'iguez et al. 2006a, b; Moya et al. 2010b).

The ADIPLS code (Christensen-Dalsgaard 2008) is one of the first and most used
adiabatic pulsation codes in the world. It is also based on a
second-order integration scheme, and it uses the Eulerian variation of
the pressure as eigenfunction. We have used the relaxation method
where the equations are solved together with a normalization
condition.  The frequencies are found by iterating on the outer
boundary condition. The calculation does not use the Cow\-ling
approximation. Finally, to compute these frequencies, we have remeshed
the model onto 2400 points and extrapolated the parameters below 0.05
$R_\odot$.

\section{Results}

Using these two codes, we have obtained the adiabatic frequency
spectrum of the Saclay equilibrium model. A complete comparison
following the work done by the ESTA group has been carried out. For
this purpose, the global characteristics of the numerical resolution
for the ADIPLS code's results have been fixed, that is: the use of the
Richar\-dson extrapolation, $P^\prime$ as eigenfunction and, $\ln r$ as
integration variable. The frequencies obtained with ADIPLS are our reference
frequencies.

On the other hand, the frequencies from GraCo have been obtained using
all of the possible options for the numerical resolution (see Moya et
al. 2008).

\begin{figure}
\includegraphics[width=80mm]{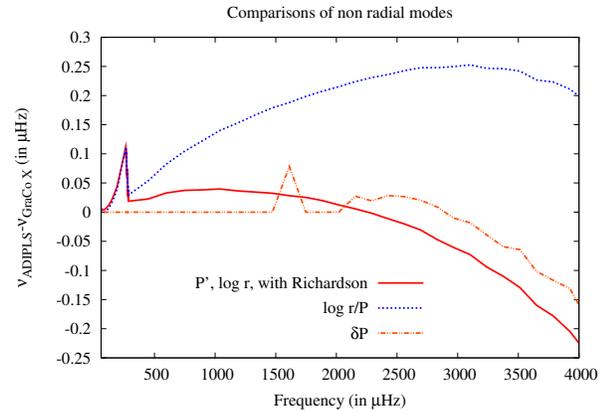}
 \caption{Overview of the differences found throughout the complete
   frequency spectrum. The ADIPLS frequencies are the
   reference. The differences obtained with all of the possibilities
   explained in the text except the case not using of the Richardson
   extrapolation are shown (see text for details).}
\end{figure}

In Figure 1, we first show,  an overview of the complete frequency
spectrum from 60 to 4000~$\mu$Hz. The differences for all the options
except the absence of the Richardson extrapolation are presented in
this comparison. If the Richardson extrapolation is not used, we reach
differences up to 10 $\mu$Hz for the largest frequencies. In this
figure, the main differences are observed when a different integration
variable is used. Nevertheless, this last choice and the rest of the
non-physical choices show differences in the range $[-0.2,
  0.3]~\mu$Hz. They are similar to those obtained by the ESTA group using a
2000 mesh-point model.

Figure 2 displays the results of the comparisons in the $g$-mode
region from 60 to 140~$\mu$Hz for the modes $\ell=1$ and 2, covering
the radial orders in the ranges n=[4,10] for $\ell=1$ and n=[7,18] for
$\ell=2$.

In this figure we see that:

\begin{itemize}

\item[i)] The comparison of frequencies for the modes $\ell =1$ and
  $\ell =2$ provides similar results.

\item[ii)] When both codes use the same configuration for the
  numerical resolution, the differences found are in the range
  $[-0.02,0.02]~\mu$Hz. These values are much lower than the
  observational accuracy. The reason of these differences must be
  searched in the use or not of the re-meshing. ADIPLS frequencies
  have been obtained using a re-mesh ``adapted'' to modes mainly
  propagating in the stellar interior, and GraCo does not have this
  option.

\item[iii)] When the Richardson extrapolation is not used, the
  differences found are in the range $[0.01,0.08]~\mu$Hz, which can
  be up to four times bigger than the other comparisons.

\item[iv)] The rest of the possible choices for the numerical
  resolution provide differences in the range
  $[-0.02,0.02]~\mu$Hz. This is similar to the range obtained when the
  same configuration is used in both codes.

\end{itemize}

\begin{figure} 
\includegraphics[width=80mm]{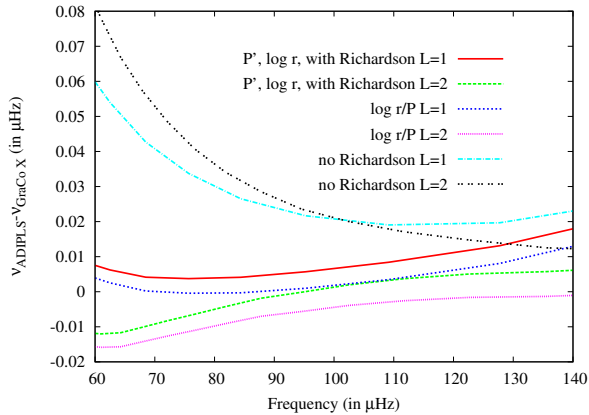}
\caption{Differences between the $g$-mode frequencies from the GraCo
  code with different options and the reference frequencies from the
  ADIPLS code as a function of the ADIPLS frequencies in the range
  $[60,140]~\mu$Hz.}
\end{figure}

\section{Conclusions}

The search for $g$ modes has been a long quest as they would be the
best probes of the solar core, thus representing a huge potential to
better constrain its structure and dynamics. Up to now, a few
candidates have been detected and recently the global properties of
dipole $g$ modes have been detected with more than 99\% confidence
level. The next step in the search for individual $g$ modes would
consist in being guided by the theoretical predictions of their
frequencies obtained with an oscillation code for a given solar model.

This is the reason why the accuracy of the frequencies calculated with
numerical algorithms is important. In this paper we have taken the
advantage of the previous studies of the ESTA group and we have tested
the accuracy of the equilibrium model used for the search of $g$ modes
in the Sun under changes of methodology in the numerical integration
of the pulsational equations. A model based on the Saclay-Seismic model has been used
as an input to the pulsational codes ADIPLS and GraCo. Two comparisons
have been studied: $i$) an overview of the differences obtained along
the complete frequency spectrum and, $ii$) a especial analysis of the
$g$-mode region $[60,140]~\mu$Hz.

This first comparison has shown that among the different methodologies
in several zones of the spectrum, the present equilibrium model
provides differences of the order of 0.1~$\mu$Hz. This also happens
when the same choices are used for both codes. This means that we need
a larger number of mesh points if we want to accurately fit the
observed frequencies in these regions.

On the other hand, the $g$-mode region presents an accuracy of the
order of $\pm~0.02~\mu$Hz for any methodology choice when the
Richardson extrapolation is used which is much better than the
uncertainties given by the physical prescriptions used in the
models. This study has shown that, if we want to pursue a search
related to observed $g$ modes in this region and based on theoretical
models, the numerical accuracy of the Saclay-Seismic model and the use
of either ADIPLS or GraCo codes, in terms of number of mesh points and
numerical accuracy of sensitive quantities, are enough. Thus, such a
guided search will be mainly sensitive to uncertainties coming from
the physical inputs of the models.

\acknowledgements {The authors want to thank
  J. Christensen-Da\-ls\-gaard who provided us the ADIPACK
  code. This work has been partially supported by the CNES/GOLF grant
  at the Service d'As\-tro\-phy\-sique (CEA/Saclay).}



\begin{thebibliography}{}

\bibitem{} Appourchaux, T., Belkacem, K., Broomhall, A.M., et al.: 2010, ARA\&A~18, 197 

\bibitem{} Asplund, M., Grevesse, N., Sauval, A. J.: 2005, in ASP Conf. Ser. 336, ed. T. G. Barnes III \& F. N. Bash (San Francisco: ASP), 25

\bibitem{} Broomhall, A.~M., Chaplin, W.~J., Elsworth, Y., Appourchaux, T.: 2007, MNRAS~379, 2

\bibitem{} Chaplin, W.~J., Christensen-Dalsgaard, J., Elsworth, Y., et al.: 1999, MNRAS~308, 405 

\bibitem{} Christensen-Dalsgaard, J.: 2008, Ap\&SS 316, 113 

\bibitem{} Couvidat, S., Turck-Chi{\`e}ze, S.,  Kosovichev, A.~G.: 2003, ApJ 599, 1434

\bibitem{} Gabriel, A.~H., Baudin, F., Boumier, P., et al.: 2002, A\&A 390, 1119 

\bibitem{} Garc{\'{\i}}a, R.~A.: 2010, Highlights of Astronomy~15, XXVIth IAU General Assembly, I.F. Corbett ed., in pess

\bibitem{} Garc{\'{\i}}a, R.~A., Corbard, T., Chaplin, W.J., et al.: 2004, Sol. Phys.~220, 269 

\bibitem{} Garc{\'{\i}}a, R.~A., Jim\'enez, A., Mathur, S., et al.: 2008a, AN~329, 476

\bibitem{} Garc{\'{\i}}a, R.~A., Mathur, S., Ballot, J.: 2008b, Sol. Phys.~251, 135

\bibitem{} Garc{\'{\i}}a, R.~A., Mathur, S., Ballot, J., et al.: 2008c, Sol. Phys.~251, 119

\bibitem{} Garc{\'{\i}}a, R.~A., R\'egulo, C., Turck-Chi\`eze, S., et al.: 2001, Sol. Phys.~200, 361


\bibitem{} Garc{\'{\i}}a, R.~A., Turck-Chi{\`e}ze, S., Jim{\'e}nez-Reyes, S.~J., Ballot, J., Pall{\'e}, P.~L., Eff-Darwich, A., Mathur, S., Provost, J.: 2007, Science~316, 1591

\bibitem{} Grevesse, N., Noels, A., Sauval, A. J.: 1993, A\&A~271, 587

\bibitem{} Hill, H., Froehlich, C., Gabriel, M., Kotov, V.~A.:
  1991, Solar interior and atmosphere (A92-36201 14-92).~Tucson, AZ,
  University of Arizona Press, 562, 562

\bibitem{} Jim{\'e}nez, A., Garc{\'{\i}}a, R.~A.: 2009, ApJS 184, 288 

\bibitem{} Lebreton, Y., Monteiro, M.~J.~P.~F.~G., Montalb\'an, J., et al.: 2008, Ap\&SS 316, 1 

\bibitem{} Mathis, S., Palacios, A.,  Zahn, J.-P.: 2004, A\&A 425, 243 

\bibitem{} Mathis, S.,  Zahn, J.-P.: 2004, A\&A 425, 229 

\bibitem{} Mathis, S.,  Zahn, J.-P.: 2005, A\&A 440, 653 

\bibitem{} Mathur, S., Eff-Darwich, A., Garc{\'{\i}}a, R.~A., Turck-Chi{\`e}ze, S.: 2008, A\&A 484, 517

\bibitem{} Mathur, S., Garc{\'{\i}}a, R.~A., Eff-Darwich, A.: 2010, Magnetic Coupling between the Interior and Atmosphere of the Sun, doi:10.1007/978-3-642-02859-5\_32

\bibitem{} Mathur, S., Turck-Chi{\`e}ze, S., Couvidat, S.,  Garc{\'{\i}}a, R.~A.: 2007, ApJ 668, 594

\bibitem{} Monteiro, M.~J.~P.~F.~G., et al.: 2006, ESA Special
  Publication 1306, 363

\bibitem{} Morel, P.: 1997, A\&AS 124, 597 

\bibitem{} Moya, A., Christensen-Dalsgaard, J., Charpinet, S., et al.: 2008, Ap\&SS~316, 231

\bibitem{} Moya, A., Garrido, R., Dupret, M.~A.: 2004, A\&A 414, 1081

\bibitem{} Moya, A., Garrido, R.: 2008, Ap\&SS 316, 129

\bibitem{} Moya, A., Mathur, S., Garc\'ia, R.~A: 2010a, Sol. Phys. in press, arXiv:1002.2132

\bibitem{} Moya, A., Amado, P.~J., Barrado, D., Garc{\'{\i}}a
  Hern{\'a}ndez, A., Aberasturi, M., Montesinos, B., \& Aceituno,
  F.: 2010b, accepted in MNRAS, arXiv:1003.3340

\bibitem{} Pall\'e, P.~L.: 1991, Advances in Space Research 11, 29

\bibitem{} Rodr{\'{\i}}guez, E. et al.: 2006a, A\&A 450, 715

\bibitem{} Rodr{\'{\i}}guez, E. et al.: 2006b, A\&A 456, 261

\bibitem{} Thompson, M.~J., Christensen-Dalsgaard, J., Miesch, M.~S.,
  \& Toomre, J.: 2003, ARA\&A 41, 599

\bibitem{} Turck-Chi{\`e}ze, S., et al.: 2001, AJ Letters 555, L69

\bibitem{} Turck-Chi{\`e}ze, S., Garc\'\i a, R.A., Couvidat, S., et al.: 2004, ApJ 604, 455 

\bibitem{} Zaatri, A., Provost, J., Berthomieu, G., Morel, P., \& Corbard, T.: 2007, A\&A~469, 1145

\end{thebibliography}
\end{document}